\documentclass[10pt, conference, compsocconf]{IEEEtran}
\usepackage{amsmath}
\usepackage{microtype}
\usepackage{algorithm,algpseudocode}
\usepackage[style=ieee,url=false,doi=false,isbn=false,sorting=none]{biblatex}
\usepackage{filecontents}
\usepackage{pgfplots,pgfplotstable,booktabs}

\title{Fast selection of N-2 contingencies for online security assessment}

\author{\IEEEauthorblockN{P.~A. Kaplunovich, K.~S. Turitsyn}
\IEEEauthorblockA{Department of Mechanical Engineering\\
Massachusetts Institute of Technology\\
Cambridge, 02139, USA\\
Email: pekap@mit.edu, turitsyn@mit.edu}}

\begin{document}
\maketitle
\graphicspath{{images/}}

\begin{center}
	{\large\bfseries Abstract}
\end{center}
\textit{We propose a novel algorithm for selection of dangerous N-2 contingencies associated with line or generator failures. The algorithm is based on iterative filtering of the set of all possible double contingencies. It is certified to identify all the dangerous contingencies, and has the complexity comparable to the N-1 contingency screening. Tests performed on realistic model of Polish power grid with about 3000 buses show that only two iterations of algorithm allow one to certify the safety of 99.9\% of all double contingencies, leaving only 0.1\% of the most dangerous ones for direct analysis.}

\section{Introduction}

North American Electric Reliability Corporation requires system operators to maintain security of the power grid and satisfy the $N-1$ contingency criterion. The consequences of single component failure contingency should be limited to a single circuit \cite{NERC2009}. However, even if the power grid is protected against single element contingency, they are still vulnerable to the events which involve multiple component outage or failures: $N-k$ contingencies. This problem will likely become even more important in coming years as the demand growth outpaces the introduction of new generation and transmission capacity, the non-dispatchable generation becomes more popular and energy markets are getting less regulated.  The problem of contingency screening for $k>1$ is computationally hard, and is practically intractable even for modest values of $N$ and k.

Multiple approaches have been proposed to address the problem of $N-k$ contingency screening problem complexity. The classical approaches involve ranking and selection methods \cite{Ejebe1979,Ejebe1988,Mikolinnas1981,Irisarri1981,Enns1982,Stott1985} that rank outage configurations due to heuristic index, new branch of such approaches investigate contingencies basing on linear outage distribution factors \cite{Davis2011,Davis2009}. There are number of approaches based on network topology analysis \cite{Chen2005,Guler2007,Dosano2009}, nonlinear optimization \cite{Mori2001,Eppstein2011,Donde2008} and others. Despite the fact that almost all researches were made concerning line contingencies, lines tripping is not the only mechanism of outages there are many of them, such as voltage collapse, relay failures, operator failure and generators tripping (on which we are focused in this work), to name of few  \cite{Vaiman2012}. It is quite common to simulate contingencies with DC power flow simplifications, however there are also investigations based on AC power flow models \cite{Chen2011,Parmer2011}, in this work we stay in DC model to avoid difficulty associated with modeling voltage collapse in steady-state.

Recently, we have proposed and alternative approach \cite{Turitsyn} that unlike the most of the common heuristic techniques is mathematically certified to identify all the dangerous $N-2$ contingencies, while maintaining the number of operations at very low level, comparable to the total number of dangerous $N-2$ contingencies. For a model system of Polish power grid with about $N\sim3000$ buses the approach decreased the total number of pairs that need to be analyzed by a factor of $2000$ at the expense of an additional overhead of $O(N^2)$ operations, comparable to the complexity of the classical $N-1$ contingency screening. In our first presentation of this approach \cite{Turitsyn} we have considered only contingencies associated with individual power line tripping events and ignored the islanding scenarios. In this proceedings we extend the approach to incorporate the generator failure events, and provide more detailed discussion of the contingency selection and certification ideas.

\section{Fast solution of contingency selection problem}

The contingency selection problem in its general form can be formulated as follows. Given a large power system with multiple elements we can identify the set of one-element contingencies ${\cal C}_1$ corresponding to configurations where one element has failed. The set of $2$-element contingencies can be formed as a Cartesian product of ${\cal C}_2 = {\cal C}_1\times {\cal C}_1$ representing the pairs of failed elements. Similarly the construction can be extended to $k$ contingencies. Every real power system is also characterized by a set of constraints ${\cal F}$, typically associated with power flows or voltage levels on individual elements.

The goal of the contingency selection problem is to identify the subset of dangerous $k$-element contingencies that violate at least one of the constraints. The complexity of the brute force approach that enumerates all the contingencies is given by $O(N^k P) + O(N^k M)$ where $N$ is the number of one-element contingencies, $P$ is the number of operation required to solve load flow problem, and $M$ is the number of constraints, i.e. the size of the set ${\cal F}$. Clearly, this approach is not feasible for real life power grids.

The natural question is whether it is possible to reduce the number of operations required to select the contingencies. Although this is not possible for arbitrary systems, one can expect more efficient algorithms to exists in actual power systems. In power systems the structure of the network is such that failures of most components produce a significant effect only in small number of constrained components. Intelligent exploitation of this property may be used for dramatic reduction of the number of contingencies that need to be analyzed. This idea has been explored in the number of recent papers \cite{Hines,Buldyrev2010}, where the notion of dual graph was introduced to represent the influence of lines on each other.

The key idea of our approach builds upon the concept of response function $\Delta^\alpha_x$ that represents the effect of contingency $\alpha \in {\cal C}$ on the power flow solutions through constrained element $x \in {\cal F}$. The concept of response function is closely related to the line outage distribution factor extensively studied in the power systems literature, and generalizes it for more general class of contingencies and constraints. For instance response functions can be used to represent the effect of generator failure on the voltage level on other generator buses, or on the phase difference on a power line.

Our algorithm is based on two key ideas. First, we argue that it is possible to relate the response functions $\Delta^{(\alpha,\beta)}_x$ of double contingencies to the response functions of individual elements $\Delta^{\alpha}_x,$  and $\Delta^{\beta}_x$. Second, using this relation one can efficiently bound the overall response for the majority of double contingencies, and certify their ``safety'' without analyzing each one of them explicitly. Combination of these two ideas leads to an effective algorithm that reduces the overall computation time by several orders of magnitude, but is guaranteed to find all the dangerous double contingencies. While the details of the algorithms are explained in Section \ref{generators} and our recent publication \cite{Turitsyn}, below we provide qualitative explanation of the key techniques used in response function analysis and contingency filtering.

\begin{figure}[ht]
    \centering{
    \includegraphics[width=5cm]{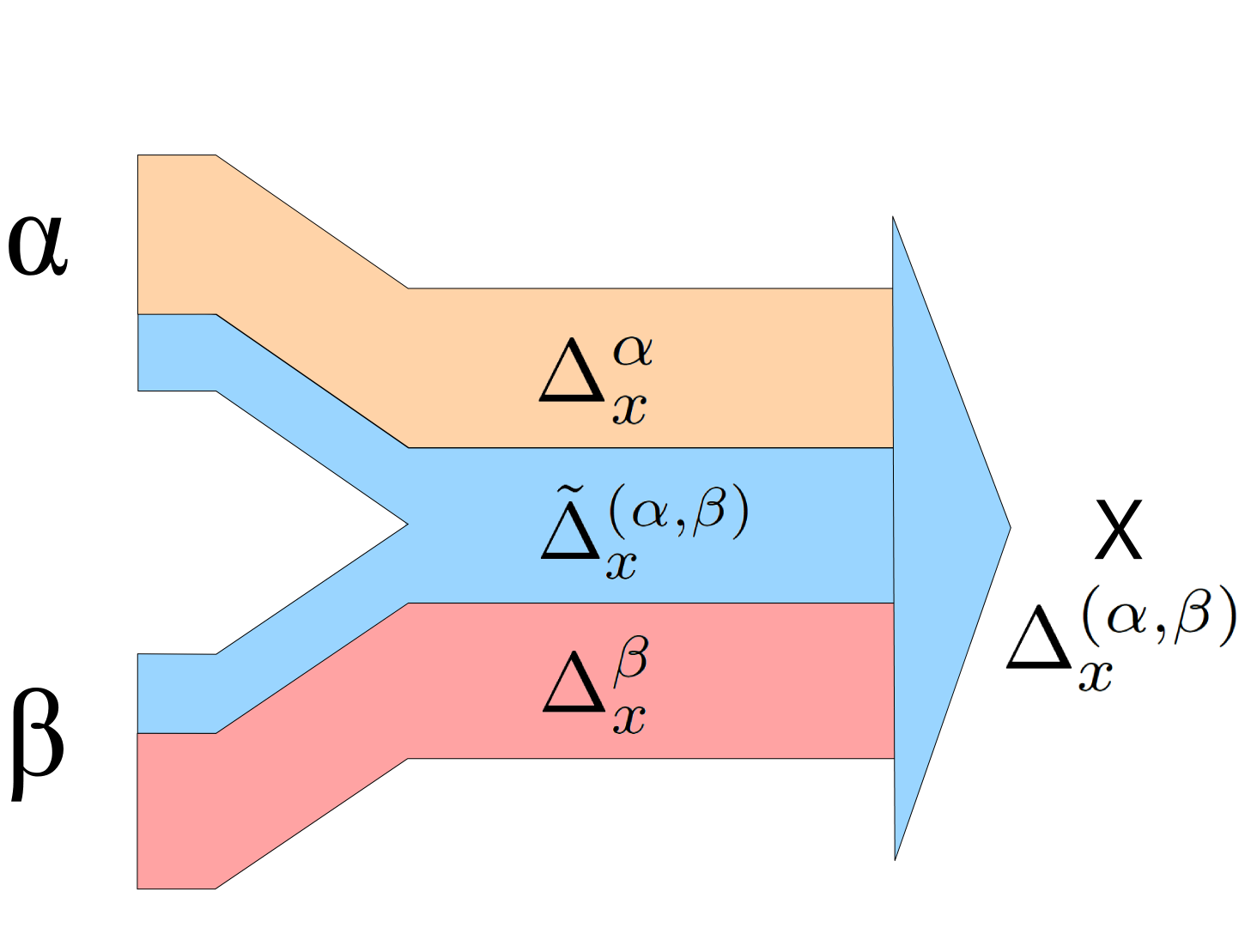}}
    \label{infl}
	\caption{Influence of the pair of contingency elements $(\alpha,\beta)$ on the constraint element $x$ through single tripping influences}
\end{figure}

The effect of double contingency $(\alpha,\beta)$ on the constrained element $x$ represented by the response function $\Delta^{(\alpha,\beta)}_x$ can be formally decomposed as
\begin{equation}
\Delta^{(\alpha,\beta)}_x = \Delta^{\alpha}_x + \Delta^{\beta}_x + \tilde{\Delta}^{(\alpha,\beta)}_x
\end{equation}
where the last term $\tilde{\Delta}^{(\alpha,\beta)}_x$ represents the effect of interference: tripping of the first element $\alpha$, either a line or a generator results in rearrangement of the power flows in the system, everywhere except on the element $\beta$. For this reason the consequent tripping of the second element $\beta$ has an effect different from single contingency represented by $\Delta^{\beta}_x$. For linear DC models one can find a simple closed form expression for the overall response function and the interference term $\tilde{\Delta}^{(\alpha,\beta)}_x$. In realistic power grids the terms $\Delta^{\alpha}_x,  \Delta^{\beta}_x$ are relatively small individually, as in normal situation the grid is protected against $N-1$ contingencies. Moreover, as most of the contingency pairs $\alpha,\beta$ correspond to lines that are far away from each other in any reasonable metric, their interference can be also bounded.

As will be shown in the next section, for linear models, like DC load flow it is possible to produce a global bound for the interference term $\tilde{\Delta}^{(\alpha,\beta)}_x$ that does not depend on the constraint $x$ in only $O(MN)$ operations. With this bound it is possible to certify the safety of most $(\alpha,\beta)$ pairs and prune the set of contingency candidates by their direct enumeration in only $O(N^2)$ operations.  So the overall complexity of the algorithm is limited to $O(N^2) + O(NM)$ operations which is $N$ times smaller than the brute-force approach. Further accelerations are possible with the introduction of other types of bounds for $\tilde{\Delta}^{(\alpha,\beta)}_x$ that do not depend on one of the contingency indices $\alpha$ or $\beta$. These bounds allow additional pruning, although does not improve overall complexity of the approach. We provide the necessary technical details of the algorithm in the next section

\section{Application to Generators} \label{generators}

In this section we discuss the application of the approach to the problem of generator failure induced contingencies. Our goal is to find all tripped generator pairs that result in an overload at least in one the power grid lines. In other words, the set of single element contingencies consists of all $N=N_G$ generators and the set of constraints consists of all $M=N_L$ lines. The brute-force enumeration of all possible $N-2$ contingencies and their effect on all the constraints would require at least $O(N_G^2 N_L)$ if we ignore the complexity of solving the DC load flow equations. Within our approach we reduce this number to $O(N_G N_L)$.

In order to do so, we need to represent the response function $\Delta_x^{(\alpha ,\beta)}$ for the overload of line $x$ after tripping to generators $\alpha,\beta$ and in terms of single contingency response: $\Delta_x^{(\alpha ,\beta)}=f(\Delta_x^{\alpha},\Delta_x^{\beta},\text{other parameters})$. Tripping of generator or load buses results in a loss of overall balance between total generation and consumption of power. In real life this balance is restored via the action of primary frequency response. In order to incorporate this effect into our analysis we introduce a simple dependence of the generator output on the shared scalar variable $\omega$ representing the deviation of the frequency from its nominal value. Note, that the  coefficient $c_\gamma$ can represent both the droop setting of the generator and the frequency response of the load.
\begin{equation}
	{P}_\gamma=P^0_\gamma\cdot(1+c_\gamma\omega)
\end{equation}
When the generator $\alpha$ is tripped, its original generation $P^0_\alpha$ is distributed among other generators resulting in the frequency shift $\omega^\alpha = P_\alpha^0/(P-c_\alpha P_\alpha^0)$, where $P =\sum_\gamma c_\gamma P_\gamma^0$. Similarly, when two generators are tripped initially, the frequency shift is given by $\omega^{(\alpha,\beta)} = (P_\alpha^0 + P_\beta^0)/(P-c_\alpha P_\alpha^0-c_\beta P_\beta^0)$. In the following, for the sake of simplicity we assume that $c_\gamma = 1$ for all the generators, and $c_\gamma = 0$ for all the load buses. The generalization to the more general situation is straightforward but rather bulky.

In the linear DC model the power flows are modeled via a vector of voltage phases $\theta$ that satisfies the load flow equations:
\begin{equation}
\hat B \theta = P,
\end{equation}
where $\hat B$ being the nodal susceptance matrix that can be represented as $\hat B = \hat M \hat Y\hat M^T$ with the diagonal $\hat Y$ matrix of branch susceptances, and connection matrix $M$. In this case the line flow vector $f$ is linearly related to the bus powers vector $P$ via $f = \hat \Delta P = \hat Y \hat M^T \hat B^{-1} P$. It is convenient for the further analysis to introduce the base vector of the power flows $f^0 = \hat \Delta P^0$, and the intermediate vector $g^\alpha$ given by $g^\alpha = \hat \Delta P^{0,\alpha}$, where $P^{0,\alpha}$ is a vector with a single non-zero component $P^0_\alpha$. In this case the single $\Delta_{x}^\alpha$ and double $\Delta_{x}^{(\alpha,\beta)}$ contingency response functions defined as the change of the flow through line $x$ after the corresponding contingency: $\Delta_{x}^\alpha = f_x^\alpha - f_x^0$.
\begin{align}
\Delta_{x}^\alpha=f^0 \omega^\alpha - g_x^\alpha(1+\omega^\alpha) \\
\Delta_{x}^\beta=f^0 \omega^\beta - g_x^\beta(1+\omega^\beta) \\
\Delta_{x}^{(\alpha,\beta)}=f^0 \omega^{(\alpha,\beta)} - (g_x^\alpha+g_x^\beta)(1+\omega^{(\alpha,\beta)})
\end{align}
This leads to the following expression for the double contingency response function:
\begin{equation}
	\Delta_{x}^{(\alpha,\beta)}=\frac{P-P_\alpha^0}{P-P_\alpha^0-P_\beta^0} \Delta_x^\alpha + \frac{P-P_\beta^0}{P-P_\alpha^0-P_\beta^0} \Delta_x^\beta
\end{equation}
This achieves our goal of expressing the effect of double contingency through the effect of single contingencies. Our next step is to develop an efficient algorithm to prune the set of contingencies that are certified to be safe. This safety certificate can be represented as the following condition:
\begin{equation}
	|f_x^0 + A_{\alpha \beta}\Delta_x^\alpha + A_{\beta \alpha} \Delta_x^\beta|<f_x^{max}
	\label{neq1}
\end{equation}
Where $A_{ij}=\frac{P-P_i^0}{P-P_i^0-P_j^0}$, $f_x^{max}$ - line power limit.
Using the representation (\ref{neq1}) we design the iterative pruning procedure described below. The sets of possible contingencies $(\alpha,\beta)$ and possible influence pairs $(\alpha,x)$ are repeatedly filtered with the help of the global bounds for the response function $\Delta_x^\alpha,\Delta_x^\beta$ and frequency shifts $\omega^{(\alpha,\beta)}$. The sequence of the steps needed to carry out the algorithm is outlined below:

\vspace{0.1cm}
\textbf{\textit{Iterative pruning algorithm}}

\vspace{0.1cm}

\textbf{Step 1 [$O(N_GN_L)$ computations]}
	
For each $\alpha$ in the set of generators find the bound $B_\alpha = \max_x \Delta^\alpha_x$ - the maximum effect of single generator tripping.

\vspace{0.1cm}

\textbf{Step 2 [$O(N_G^2)$ computations]}

Check the condition $A_{\alpha \beta} B_\alpha + A_{\beta \alpha} B_\beta<1$ for all $(\alpha,\beta)$. Filter out all pairs $(\alpha, \beta)$ that don't satisfy this inequality. In this step the set ${\cal C}_2$ of potentially dangerous pairs $(\alpha,\beta)$ is decreased due to the existence of the global bound $B_\alpha$ for single contingencies.

\vspace{0.1cm}

\textbf{Step 3 [$O(N_GN_L)$ computations]}

Similarly filter out the set of potential pairs $(\alpha,x)$. In this step we bound the "interference" effect. This is an optional step, that requires construction of additional bounds, details of the procedure can be found in \cite{Turitsyn}.

\vspace{0.2cm}

These steps need to be iterated until the set of double contingency candidates ${\cal C}_2$ stops changing, or its size reaches some satisfactory value. The complexity of the algorithm is $O(N_G^2+N_LN_G)$, so in essence we certify safety of a lot of triples $(\alpha,\beta,x)$ without directly checking them.

\section{Results}
We have tested and validated the algorithm using the largest publicly available grid model, which is the Polish Power Grid available in MATPOWER package \cite{Zimmerman2011}. This power grid consists of $N_G=204$ generators and $N_L=3269$ power lines each with a power constraint. The total number of generator pairs that need to be considered by the contingency screening algorithm equals to $(N_G-1)(N_G)/2=20706$. However, the total number of actually dangerous pairs is just $89$. Brute-force approach of checking all the constraints for all the generator pairs would require approximately $70$ million operations.

When the set of contingency pairs is filtered with our algorithm the total number of double contingency candidates is reduced to $189$ after the first iteration and to $114$ after the second. The overall complexity of the iteration is estimated as $O(N_G N_L)$, which corresponds to approximately $6\cdot 10^5$ operations, about hundred times smaller than the number of operations required by brute force enumeration.

\begin{table}[ht]
\begin{center}
\pgfplotstabletypeset[
	int detect,
	columns={iter,N},
	columns/iter/.style={column name=\textsc{Iteration}},
	columns/N/.style={column name=Number of pairs},
	every head row/.style={before row=\toprule,after row=\midrule},
	every last row/.style={after row=\bottomrule}
]{
iter	N
0	20706	
1	189	
2	114
3	114
}
\end{center}
\caption{Candidate set of generators pairs size evolution with algorithm progression.}
\label{gentable}
\end{table}

\newcommand{\specialcell}[2][c]{%
  \begin{tabular}[#1]{@{}c@{}}#2\end{tabular}}

  \begin{table}[hh]
\centering
\begin{tabular}{c c c c}
\toprule
Load & \specialcell{$1^{\text{st}}$ Iteration \\ of algorithm} & \specialcell{Direct \\ Enumeration} &	 \specialcell{$1^{\text{st}}$ Iteration \\ Without dangerous line}\\
\midrule
0$\%$ &	189 & 89 & - \\	
5$\%$ &	1899	& 1473 & 16\\
10$\%$ &	6270	& 5492 & 29\\
15$\%$ &	\multicolumn{2}{c}{No feasible solution}	    & 99\\	
\bottomrule
\end{tabular}
\caption{Algorithm performance on overloaded grid}
\label{loadres}
\end{table}

In order to understand how the algorithm performs in more stressed conditions we have done several experiments on the loaded Polish Grid. In all of these experiments we increased by the same percentage the consumption and generation levels to push the system closer to thermal limits. The number of dangerous double contingencies increases significantly as can be seen from the Table \ref{loadres}. However, our algorithm manages to find all of them, and the number of false candidates seems to be proportional to the total number of dangerous contingencies.

When analyzing the results we have found that most of these contingencies result in the overload of a single line close to its limit. This a natural modification of our algorithm. If we separate the most overloaded line in a separate set, the effect of all the double contingencies on this line can be tested only in $O(N_G^2)$ operations. However, as seen from the table the number of candidate pairs that can result in overload of other lines is reduced by several order of magnitudes. In the case of $10\%$ additional load our original algorithm gives $6270$ contingencies after first iteration. However if the most dangerous line is removed, only $29$ contingencies are left after first iteration. If we increase the load by $15\%$ any feasible solutions disappear.

Separation of constraint set is potentially a powerful way of improving the performance of the algorithm in the cases where the number of potentially dangerous contingencies is large. The most dangerous contingencies can be identified in a single loop over the set of generator$\to$ constraint pairs, after which the constraints are ranked in the order of the appearance frequency or some other heuristic. We have done extensive experiments with this idea, but we expect it to result in significant improvements in highly loaded systems.

Finally, we have to emphasize again, that although we have focused in this paper on the contingencies associated with generator failures, the approach can be also applied to more common line failures.
In our previous publication \cite{Turitsyn} we have carried out the simulations for line tripping events, so the set of double contingencies ${\cal C}_2$ consisted of all line pairs. Only $4$ iterations with overall complexity of $O(N_L^2)$ operations resulted in the decrease the size of candidate set size from $5341546$ to $5750$ pairs (see Table. \ref{linestable}). At the same time the total number of actual dangerous double contingencies was $524$. Although the number of false candidates in this case was large, the complexity of their direct analysis is still negligible, and comparable to the complexity of $N-1$ contingency analysis \cite{Turitsyn}.

\begin{table}[ht]
\begin{center}
\pgfplotstabletypeset[
	int detect,
	columns={iter,N},
	columns/iter/.style={column name=\textsc{Iteration}},
	columns/N/.style={column name=Number of pairs},
	every head row/.style={before row=\toprule,after row=\midrule},
	every last row/.style={after row=\bottomrule}
]{
iter	N
0	5341546	
1	17928	
2	6128	
3	5816	
4	5750	
5	5750
}
\end{center}
\caption{Candidate set size evolution with algorithm progression.}
\label{linestable}
\end{table}

One can see that the algorithm performs well for any type of contingencies. The underlying reason for its efficiency is the relative reliability of the operating point of realistic power grids. As the grids are protected against $N-1$ contingencies, most of the pair contingencies do not result in violation of any constraints. Our algorithm allows to filter them out without actually solving the power flow for every one of them. The mathematical foundation of the algorithm allows us to guarantee that all the filtered contingencies are safe, and no dangerous contingency is ever missed.


\section{Conclusions}

In conclusion, we have presented a novel contingency selection algorithm for fast identification of dangerous $N-2$ contingencies. The algorithm reduces the overall number of operation by a factor $~N$ in comparison to straightforward enumeration of all contingencies. The resulting acceleration can be very impressive for realistic large scale networks. The version of the algorithm presented in these proceedings extends our original paper \cite{Turitsyn} with the inclusion of generator failure contingencies.

In comparison to alternative heuristics designed to address the complexity of $N-k$ contingency screening problems the main advantage of our approach is that it is mathematically certified to identify all the dangerous contingencies. No dangerous contingencies can be missed, and the only drawback of the algorithm is that some of the safe contingencies are included in the final candidate set. The number of these safe contingencies was found to be small in practically relevant models, although there is no mathematical bound on their number.

At this moment our approach works only for DC power flow model, and is limited to $N-2$ contingencies. Our preliminary studies indicate that it is possible to achieve strong reductions in computation time even in $N-k$ situations. However it is still not clear whether it is possible to avoid the curse of dimensionality and change the complexity in $k$ from exponential to polynomial. Also, it is not yet clear whether this technique can be generalized to nonlinear AC power flow models as it relies heavily on the algebraic structure of linear equation solution. In our opinion the most promising ideas that have to be explored in this direction are the attempts to bound the nonlinear terms, exploiting the fact the the power systems operate in weakly nonlinear regime. These kind of bounds can be naturally incorporated in our procedure and make it work even for more realistic models.

\section{Acknowledgements}

We thank NSF and MIT/SkTech initiative for their support. KT thanks the participants of LANL smart grid seminar for the feedback on preliminary reports of these results.
The work is partially supported by the Russian Federal Targeted
Programs "S\&S-PPIR" and "I\&DPFS\&T".


\printbibliography

\end{document}